\documentclass[showpacs,preprintnumbers,amsmath,amssymb,onecolumn,floatfix,aps,pre]{revtex4}

\usepackage{graphicx}% Include figure files
\usepackage{dcolumn}
\usepackage{bm}
\usepackage{epsfig}
\usepackage{amsmath}
\usepackage{amssymb}
\newcommand{\ensemblemean}[1]{\langle\rangle}

\newcommand{\qed}{\nobreak \ifvmode \relax \else
      \ifdim\lastskip<1.5em \hskip-\lastskip
      \hskip1.5em plus0em minus0.5em \fi \nobreak
      \vrule height0.75em width0.5em depth0.25em\fi}

\begin{document}

\title{Analytical properties of horizontal visibility graphs in the Feigenbaum scenario}
% Force line breaks with \\
\author{Bartolo Luque$^1$, Lucas Lacasa$^{1*}$, Fernando J. Ballesteros$^2$, Alberto Robledo$^{3}$}
\affiliation{$^1$Dept. Matem\'{a}tica Aplicada y Estad\'{i}stica\\
                ETSI Aeron\'{a}uticos, Universidad Polit\'{e}cnica de Madrid, Spain\\
$^2$  Observatori Astron\`{o}mic,\\
                Universitat de Val\`{e}ncia, Spain\\
$^{3}$ Instituto de F\'{\i}sica y Centro de Ciencias de la Complejidad, Universidad Nacional Aut\'{o}noma de M\'{e}xico, Mexico\\
}%

\date{\today}
% It is always \today, today, % but any date may be
% explicitly specified

\email{lucas.lacasa@upm.es}


\begin{abstract}
Time series are proficiently converted into graphs via the
horizontal visibility (HV) algorithm, which prompts interest in
its capability for capturing the nature of different classes of
series in a network context. We have recently shown \cite{plos}
that dynamical systems can be studied from a novel perspective via
the use of this method. Specifically, the period-doubling and
band-splitting attractor cascades that characterize unimodal maps
transform into families of graphs that turn out to be independent
of map nonlinearity or other particulars. Here we provide an in
depth description of the HV treatment of the Feigenbaum scenario,
together with analytical derivations that relate to the degree
distributions, mean distances, clustering coefficients, etc.,
associated to the bifurcation cascades and their accumulation
points. We describe how the resultant families of graphs can be
framed into a renormalization group scheme in which fixed-point
graphs reveal their scaling properties. These fixed points are
then re-derived from an entropy optimization process defined for
the graph sets, confirming a suggested connection between
renormalization group and entropy optimization. Finally, we
provide analytical and numerical results for the graph entropy and
show that it emulates the Lyapunov exponent of the map
independently of its sign.
\end{abstract}
\maketitle%%%
%05.45.Tp   Time series analysis
%05.45.Ac   Low-dimensional chaos
%89.75.Hc   Networks and genealogical trees

\textbf{In recent years a new general framework to make time series analysis has been coined. This framework is based on the mapping of a time series into a network representation and the subsequent graph theoretical analysis of the network, offering the possibility of describing the structure of complex signals and the associated dynamical systems from a new and complementary viewpoint, and with a full set of alternative measures. Here we focus on a specific type of mapping called the horizontal visibility algorithm, and via this approach we address the specific case of the period-doubling route to chaos. We extend our preliminary results on this topic \cite{plos}, and provide a complete graph theoretical characterization of unimodal iterated maps undergoing period doubling route to chaos that, we show, evidence a universal character. Our approach allows us to visualize, classify and characterize periodic, chaotic and onset of chaos dynamics in terms of their associated networks.}

\section{Introduction}
Very recently \cite{plos}, a connection between nonlinear
dynamical systems and complex networks has been accounted for by
means of the horizontal visibility (HV) algorithm
\cite{pre,submitted}, as the latter transforms time series into
graphs. The families of trajectories generated by nonlinear
low-dimensional iterated maps conform a distinctive class of time
series. Accordingly, they make up ideal candidates to test the
capabilities of the HV algorithm for capturing meaningfully the
information contained in them, and, if so, see how these manifest
in the network central quantities. The possibility of observation
of novel properties adds to the motivation to carry on these
studies.  We have chosen to inspect first the well-known
one-dimensional (therefore dissipative) unimodal maps and their
common period-doubling route to chaos, when periodic attractors
transform into aperiodic attractors, the bifurcation cascade or
Feigenbaum scenario \cite{chaos,chaos2}. This route to chaos
appears an infinite number of times amongst the family of
attractors generated by unimodal maps within the windows of
periodic attractors that interrupt sections of chaotic attractors.
In the opposite direction, a route out of chaos accompanies each
period-doubling cascade by a chaotic band-splitting cascade, and
their shared bifurcation accumulation points form transitions
between order and chaos that possess universal properties
\cite{chaos,chaos2,steve}. Low-dimensional dynamics benefits from
added interest as systems with many degrees of freedom relevant to
various problems in physics and elsewhere are known to undergo a
drastic simplification and display this type of dynamics
\cite{Strogatz}.

There is a growing number of methods designed to transform series
into networks, involving concepts such as recurrence in phase space \cite{zhang1,review_grafos} or Markov processes \cite{amaral}
to cite a few, and our approach forms part of this enterprise \cite{pnas}. Once a time series
is converted into a network the interest lies in the observation
of the characteristic properties of dynamical systems in a
different environment. And to achieve this it is necessary to use
the characteristic tools of network analysis
\cite{stro1,redes,redes2,redes3,bollobas}. The family of
visibility algorithms has been successful in obtaining information
relevant to the description of fractal behavior \cite{epl} or to
the distinction between random and chaotic series
\cite{submitted}. Here we detail the Feigenbaum scenario as seen
through the HV formalism by providing a complete description of
its associated set of graphs. These graphs represent the time
evolution of all trajectories that take place within the
attractors of unimodal maps. The outline of the presentation is
the following: We start in Section II by recalling the
construction of an HV graph from a time series and deduce general
expressions for the mean degree and distance when the series is
periodic. We advance a visual illustration of the graphs and their
location in the Feigenbaum diagram. In Section III we focus on the
period-doubling cascade and derive a simple closed-form expression
for the degree distribution of periodic attractor graphs and their
accumulation point. We obtain from the latter the mean degree, the
variance and the clustering coefficient. In Section IV we center
on the reverse bifurcation cascade of chaotic-band attractors and
derive the expression for the degree distribution and the mean
degree. As the number of bands increases there is a growing
similarity with the same quantities for the period-doubling
cascade, since, as shown, the contribution from chaotic motion is
confined only to the shrinking top band. The properties of graphs
stemming from both chaotic-band attractors and windows of
periodicity are derived with help of the self-affine properties of
the bifurcation cascades. In Section V we describe a
renormalization group (RG) transformation, equivalent to the
original functional composition RG transformation, but specially
designed for the Feigenbaum graphs, that leads to a set of fixed
point graphs that further explain, and give unity to, the two
previous sections. In Section VI we turn attention to the entropy
associated to the degree distributions and find that under
optimization we recover the RG fixed points. Finally we compare
the behavior of this entropy as we move along the bifurcation
cascades and notice that this quantity follows closely the
variation of the map's Lyapunov exponent, pointing out to a
property reminiscent of the Pesin equality but suitable for both
periodic and chaotic graphs. In Section VII we summarize our
results. A brief preliminary account
of the contents of this paper is given in \cite{plos}.\\

\section{Feigenbaum graphs}
\begin{figure}[!htb]
\centering
\includegraphics[width=0.6\textwidth]{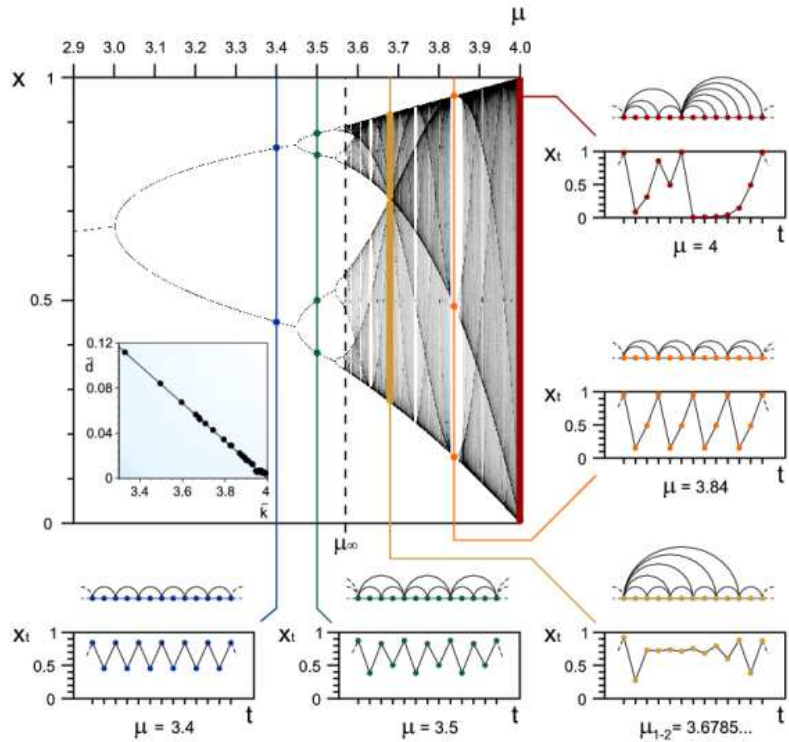}
\caption{Feigenbaum graphs from the Logistic map $x_{t + 1}  =
f(x_t)=\mu x_t (1 - x_t )$. The main figure portrays the family of
attractors of the Logistic map and indicates a transition from
periodic to chaotic behavior at $\mu_\infty = 3.569946...$ through
period-doubling bifurcations. For $\mu\geq\mu_\infty$ the figure
shows merging of chaotic-band attractors where aperiodic behavior
appears interrupted by windows that, when entered from their
left-hand side, display periodic motion of period $T=m\cdot2^0$
with $m>1$ (for $\mu<\mu_{\infty}$, $m=1$) that subsequently
develops into $m$ period-doubling cascades with new accumulation
points $\mu_\infty(m)$. Each accumulation point $\mu_\infty(m)$ is
in turn the limit of a chaotic-band reverse bifurcation cascade
with $m$ initial chaotic bands, reminiscent of the self-affine
structure of the entire diagram. All unimodal maps exhibit a
period-doubling route to chaos with universal asymptotic scaling
ratios between successive bifurcations that depend only on the
order of the nonlinearity of the map \cite{Feigenbaum}, the
Logistic map belongs to the quadratic case. Adjoining the main
figure, we show time series and their associated Feigenbaum graphs
according to the HV mapping criterion for several values of $\mu$
where the map evidences both regular and chaotic behavior (see the
text). \textit{Inset:} Numerical values of the mean normalized
distance $\bar d$ as a function of mean degree $\bar k$ of the
Feigenbaum graphs for $3<\mu<4$ (associated to time series of
$1500$ data after a transient and a step $\delta \mu=0.05$), in
good agreement with the theoretical linear relation (see the
text).} \label{figintro}
\end{figure}
The HV graph \cite{pre} associated with a given time series
$\{x_i\}_{i=1,. . .,N}$ of $N$ real data is constructed as
follows: First, a node $i$ is assigned to each datum $x_i$, and
then two nodes $i$ and $j$ are connected if the corresponding data
fulfill the criterion $x_i,x_j > x_n$ for all $n$  such that $i <
n < j$. Let us now focus on the Logistic map \cite{chaos} defined
by the quadratic difference equation $x_{t + 1}  = f(x_t)=\mu x_t
(1 - x_t )$ where $x_t \in [0,1]$ and the control parameter $\mu
\in {\rm [0,4]}$. According to the HV algorithm, a time series
generated by the Logistic map for a specific value of its control parameter $\mu$ (after
an initial transient of approach to the attractor) is converted
into a Feigenbaum graph (see figure \ref{figintro}). Notice that
this is a well-defined subclass of HV graphs where consecutive
nodes of degree $k=2$, that is, consecutive data with the same
value, do not appear, what is actually the case for series
extracted from maps (besides the trivial case of a constant
series). Also, as proven in
\cite{simone}, an HV graph is, by
construction, a planar graph, that is, it has a diagram
representation in which any pair of links intersect only at their
endpoints. Moreover, an HV graph is also outerplanar: each node
contacts the infinite face, where a face is a bounded region of a
planar graph and the infinite face is its outer region. In what follows we take advantage of these facts and outline some generic properties of
Feigenbaum graphs.\\
%While for a period $T$ there are in principle
%several possible periodic orbits, and therefore the set of
%associated Feigenbaum graphs is degenerate, it can be proved that
%the mean degree $\bar k(T)$ and normalized mean distance $\bar
%5d(T)$ of all these Feigenbaum graphs fulfill $\bar
%k(T)=4(1-\frac{1}{2T})$ and $\bar d(T)=\frac{1}{3T}$ respectively,
%yielding a linear relation $\bar d(\bar k)=(4-\bar k)/6$ that is
%corroborated in the inset of
%figure \ref{figintro}. In what follows we provide proofs for these results.\\
%\subsection{Mean degree and normalized distance of Feigenbaum
%graphs}

\noindent \textbf{\emph{Mean degree $\bar{k}$.}}
\begin{figure}[h]
\centering
\includegraphics[width=0.6\textwidth]{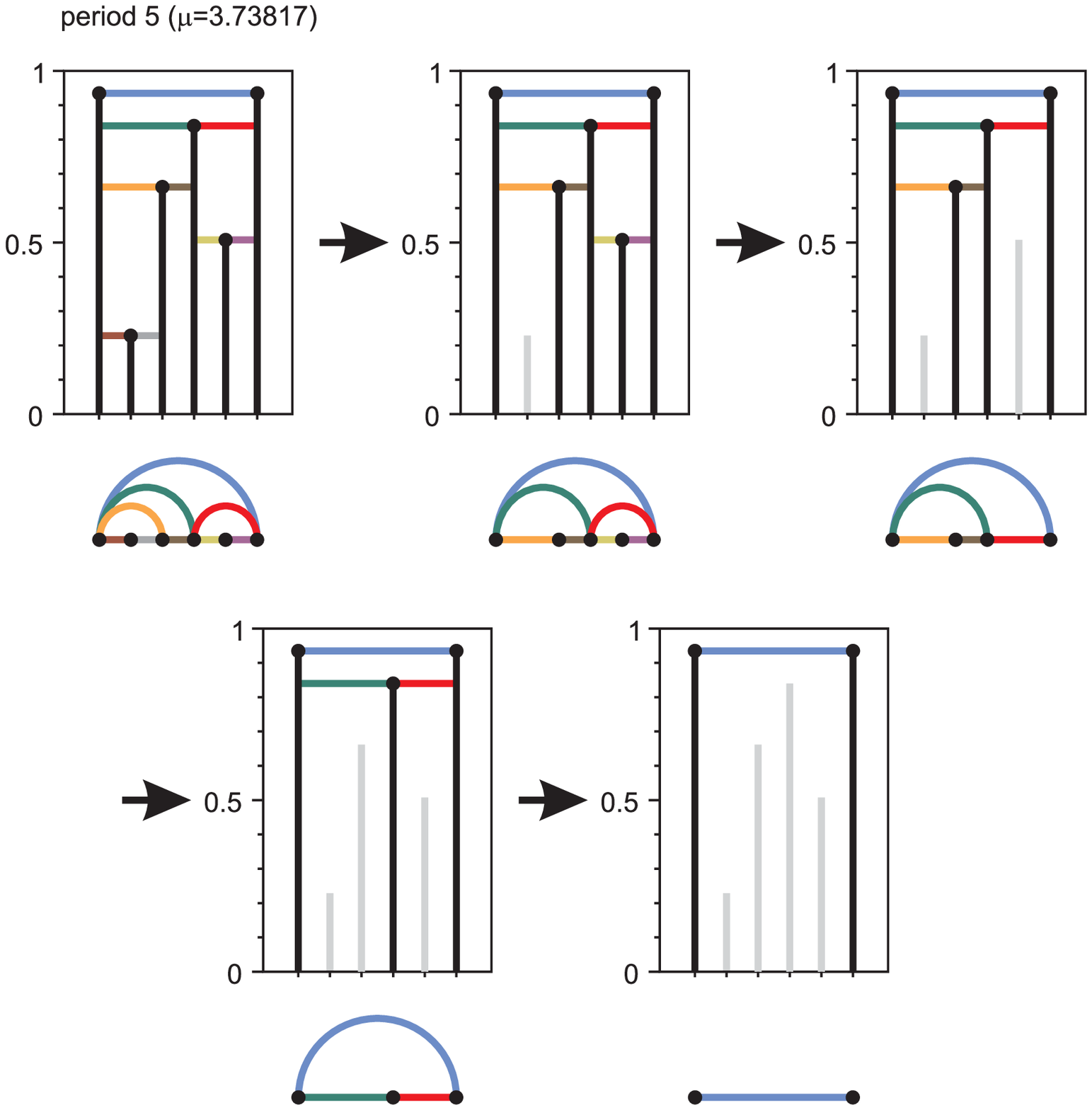}
\caption{Graphical illustration of the constructive proof of the
expression for the mean degree $\bar{k}(T)$ through consideration
of a motif extracted from a periodic series of period $T=5$.
Observe that the second data is the one with the lowest value. By
deleting it, the graph loses $2$ links. This process is iterated
three more times. A total amount of $2\cdot(5-1)=8$ links has been
deleted, independently of the inner structure of the $T=5$ motif.}
\label{ejemplo_eli}
\end{figure}
Consider a periodic orbit of period $T$. Without lack of generality, we represent the orbit as the infinite time series $%
\{...,x_{0},x_{1},...,x_{T},x_{1},x_{2},...\}$, where
$x_{0}=x_{T}$ corresponds to the largest value of the series. By
construction, the associated Feigenbaum graph consists of a
concatenation of identical motifs of $T+1$ nodes associated to the subseries $%
\{x_{0},x_{1},...,x_{T}\}$. Suppose that the motif is a graph with
$V$ links, and let $x_{i}$ be the smallest datum of the subseries
which, by construction, will have degree $k=2$ (and since no data
repetitions are allowed in the motif, $x_{i}$ will always be well
defined). Now remove this node and its two links from the motif.
The
resulting motif will have $V-2$ links and $T$ nodes. Iterate this operation $%
T-1$ times (see  figure \ref{ejemplo_eli} for a graphical
illustration of this process in a particular case with $T=5$). The
resulting graph will have only two nodes, associated with $x_{0}$
and $x_{T}$, connected by a single link, and the total number of
deleted links will be $2(T-1)$. The mean degree $\bar{k}$ of the
graph corresponds to the mean degree of the motif made of $T$
nodes (the nodes associated with $x_{0}$ and $x_{T}$ only
introduce half of their degree in the motif, what is equivalent to
an effective reduction of one node). Hence,
\begin{equation}
\bar{k}(T)\equiv 2\frac{\#\ edges}{\#\ nodes}=\frac{2(2(T-1)+1)}{T}%
\Rightarrow \bar{k}(T)=4\bigg(1-\frac{1}{2T}\bigg).  \label{th1}
\end{equation}%

\noindent The above result holds for every periodic or aperiodic
($T\rightarrow\infty$) series, independent of the deterministic
process that generates them, as the only constraint in its
derivation is that data within a period are not repeated. It
therefore includes all graphs generated by unimodal maps
irrespective of their degree of nonlinearity. Observe that the
maximum mean degree (achieved for aperiodic series) is
$\bar{k}(\infty )=4$, in agreement with previous theory (see
\cite{simone}).\\

\noindent \textbf{\emph{Normalized mean distance $\bar{d}$.}} On
the other hand, the normalized mean distance $\bar{d}$ of the
graph is defined as $\bar{d}=\bar{D}/N$, where $\bar{D}$ is the
mean distance (the average over all pairs of nodes of the smallest
path that connects each
pair) and $N$ the number of nodes. For graphs associated with periodic orbits $%
\bar{d}$ depends on $T$ (as this is the maximal amount of nodes
that can be jumped through a link), and straightforwardly gives
$\bar{d}(T)=\frac{1}{3T}$ for $N\rightarrow \infty $. Therefore,
for HV graphs $\bar{d}$ and $\bar{k}$ are linearly related by

\begin{equation}
\bar{d}(\bar{k})=\frac{1}{6}(4-\bar{k}). \label{d_k}
\end{equation}%

This latter analytical relation is checked numerically in the
inset of figure \ref{figintro}. The limiting solution
$\bar{k}\rightarrow 4$, $\bar{d}\rightarrow 0$ holds for all
aperiodic, chaotic or random series. In addition to the numerical
results shown in the inset of figure \ref{figintro} for the
specific case of the Logistic map, we have also examined the
accuracy of the latter relation for several unimodal maps, giving
perfect agreement in every case (data not shown).

%%%%%%%%%%%%%%%%%%%%%%%%%%%%%%%%%%%%%%%%
%%%%%%%%%%%%%%%%%%%%%%%%%%%%%%%%%%%%%%%%%
%%%%%%%%%%%%%%%%%%%%%%%%%%%%%%%%%%%%%%%%%%%
\section{Period-doubling route to chaos: Results}

\begin{figure}[!htb]
\centering
\includegraphics[width=0.8\textwidth]{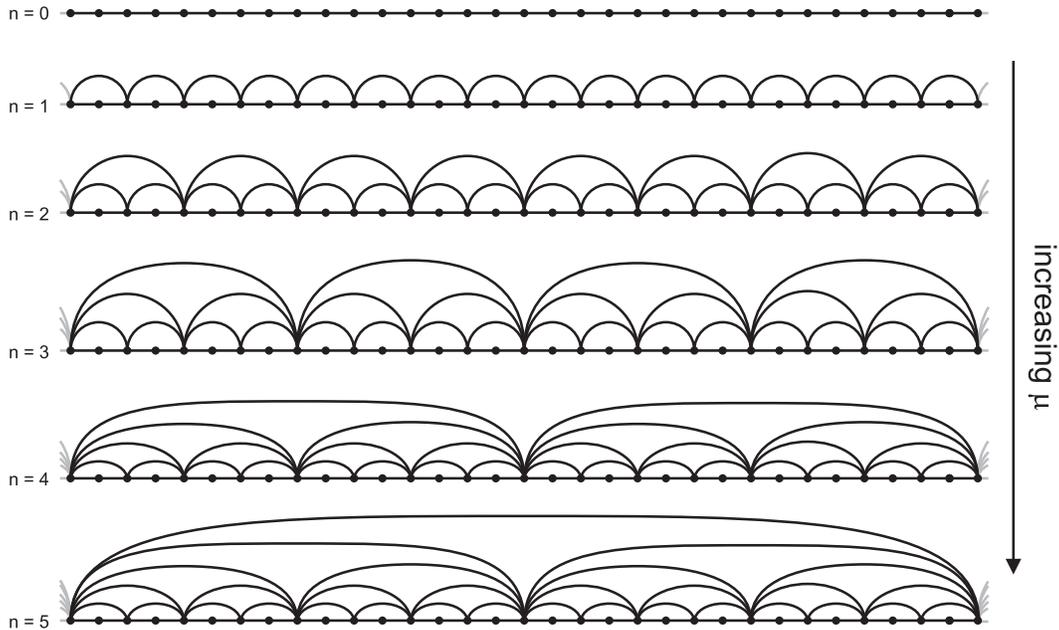}
\caption{Periodic Feigenbaum graphs for $\mu<\mu_\infty$. The
sequence of graphs associated to periodic attractors with
increasing period $T=2^n$ undergoing a period-doubling cascade.
The pattern that occurs for increasing values of the period is
related to the universal ordering with which an orbit visits the
points of the attractor. Observe that the hierarchical
self-similarity of these graphs requires that the graph for $n-1$
is a subgraph of that for $n$.} \label{grafos de feigenbaum1}
\end{figure}

\subsection{Order of visits of stable branches and chaotic bands}
\begin{figure}[h]
\centering
\includegraphics[width=0.70\textwidth]{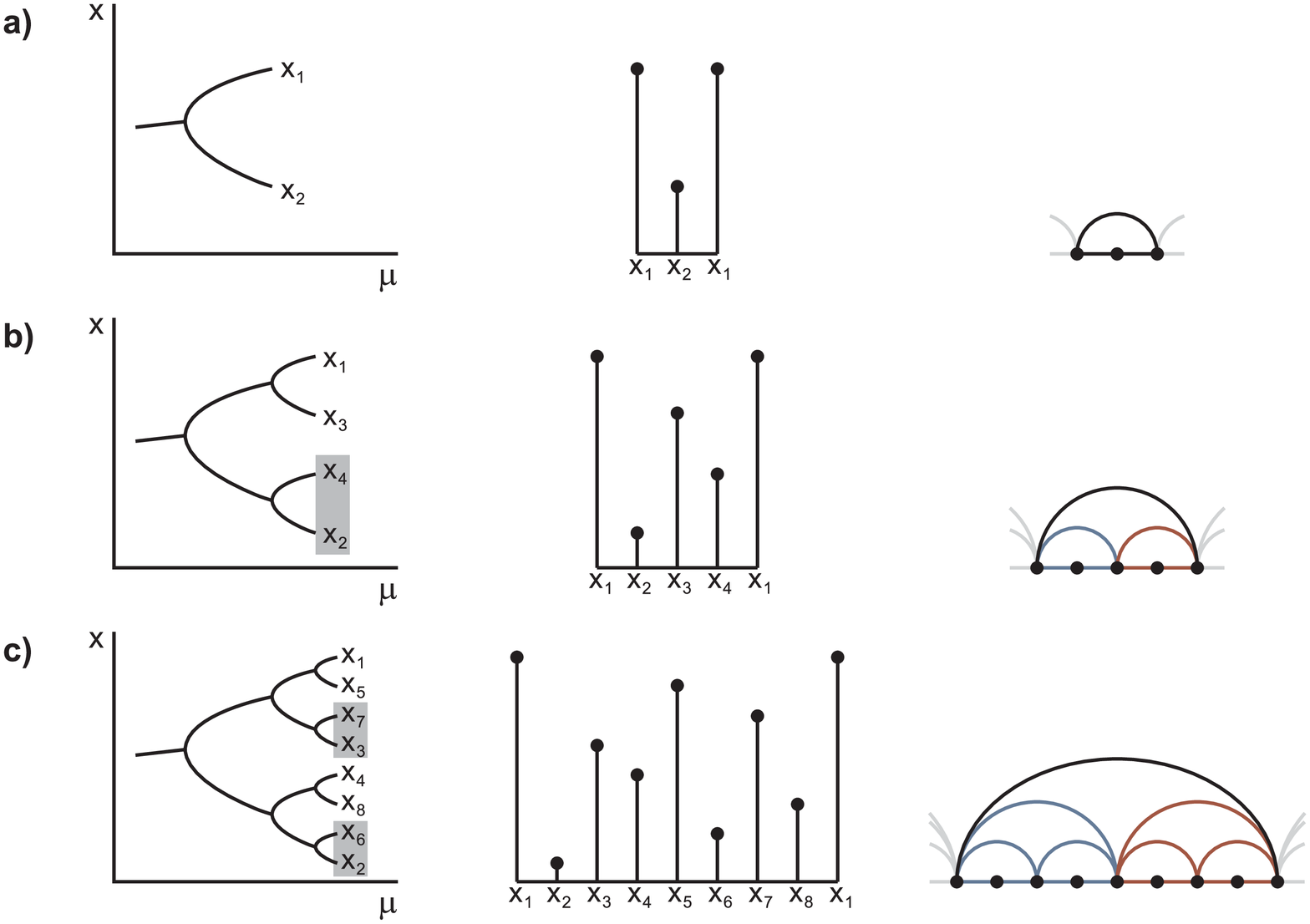}
\caption{{\protect\small {Graphical illustration that explains how
the order of visits to the stable branches of the map induces the
structure of the Feigenbaum graphs all along the period-doubling
bifurcation cascade ($\mu<\mu_\infty$).}}} \label{order}
\end{figure}

A deep-seated feature of the period-doubling cascade is that the
order in which the positions of a periodic attractor are visited
is universal \cite{schroeder}. That is, the visiting
 order of the positions $\{x_{i}\}$ (where the subindex $i$ denotes the iteration
  time) of a periodic attractor along the period-doubling route to chaos
is the same for all unimodal maps \cite{schroeder}. This ordering
turns out to be a decisive property in the derivation of the
structure of the Feigenbaum graphs (see figure \ref{grafos de
feigenbaum1} where we plot the graphs for a family of attractors
of increasing period $T=2^n$, that is, for increasing values of
$\mu<\mu_{\infty}$). Here we describe the rule that such ordering
follows for orbits of period $T=2^{n}$, and how this in turn
induces the structure of the associated Feigenbaum graphs. This is
illustrated graphically in figure \ref{order}.\\

Consider the first period-doubling bifurcation that produces
attractors with period $T=2^{1}$ and for which repeated jumps are
observed between two positions in time, $x_{1}$, $x_{2}$, $x_{1}$,
..., with $x_{1}>x_{2}$ (figure \ref{order}.a). Without lack of
generality label $x_{1} $ as the largest data. This series is
transformed into a Feigenbaum graph made up of a concatenation of
a root motif of $3$ nodes, where by construction the inner node is
associated with datum $x_{2}$. As the family of attractors reaches
the next period-doubling bifurcation, each point $x_{i}$
of the period-2 attractor splits into two new stable `offspring' positions: $%
x_{i}$ and $x_{i+T}$, and the visiting order is such that $%
x_{1}>x_{3}>x_{4}>x_{2}$ (figure \ref{order}.b). This ordering is
reminiscent of the $T=2^{1}$ orbit that was present before the
bifurcation (namely, the orbit returns to a neighborhood of the
point after a journey along the attractor). In particular, this
means that the second
largest value $x_{3}$ (the offspring of $x_{1}$) is visited only after a $%
T=2^{1}$ journey, that is, in the middle of the $T=2^{2}$ periodic
orbit. Observe also that the bottom pair of offspring positions
appears inverted (grey box in the figure). The corresponding
Feigenbaum graph is a concatenation of the $T=2^{1}$ motif (blue
and red portions in the figure) linked by the largest node
$x_{1}$, which repeats after $T=2^{2}$ iteration times. Blue and
red portions in the graph are equivalent since the orbit follows
the same pattern of visits across the stable branches (each
portion according to a given offspring). The same procedure can be
iterated for increasing period-doubling bifurcations (see figure
\ref{order}.c), leading to Feigenbaum
graphs which are progressively self-similar and become so in the limit $%
n\rightarrow \infty $. Summing up, the period-doubling bifurcation
of an orbit of period $T=2^{n}$ generates two identical copies of
the $T=2^{n}$ root motif of the graph that are now concatenated by
the node associated to datum $x_{1+2^{n}}$ and linked by the
bounding nodes $x_{1}$, and this in turn is the root motif of the
$T=2^{n+1}$ Feigenbaum graph. In the following section we will
take advantage of this structure to analytically derive several
topological properties of the Feigenbaum graphs along the
period-doubling cascade.

\subsection{Topological properties of Feigenbaum graphs along the period-doubling
cascade}
\subsubsection{Degree distribution $P(n,k)$}
\begin{figure}[h]
\centering
\includegraphics[width=0.4\textwidth]{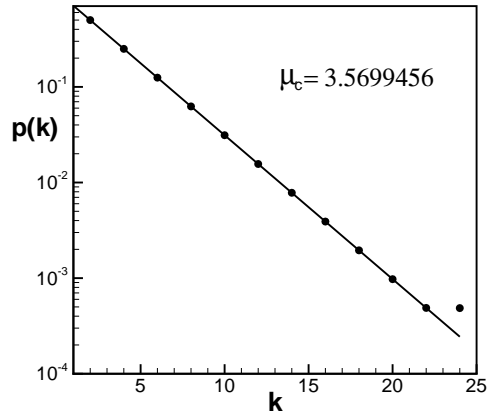}
\caption{{\protect\small {Dots: Semi-log plot of the degree
distribution of a Feigenbaum graph associated with a time series
of $10^{6}$ data extracted from a Logistic map at the onset of
chaos $\protect\mu _{\infty}=3.5699456...$. The straight line
corresponds to equation \protect\ref{degree_p_1_critico}, in
agreement with the numerical calculation (the deviation for large
values of the degree are due to finite size effects).}}}
\label{distkmucritica}
\end{figure}
\noindent The above-described order of visits generates a
hierarchy of self-similar Feigenbaum graphs along the
period-doubling bifurcation cascade. The degree distribution of a
graph is defined as a discrete probability distribution that
expresses the probability of finding a node with degree $k$
\cite{redes,redes2,redes3}. By construction, the degree
distribution of a Feigenbaum graph for a series of period $T=2^{n}$, $%
n=0,1,2,...,$ is
\begin{eqnarray}
&&P(n,k)=\left( {\frac{1}{2}}\right) ^{k/2},\quad k=2,4,6,...,2n,
\notag
\label{degree_p_n} \\
&&P(n,k)=\left( {\frac{1}{2}}\right) ^{n},\quad k=2(n+1),  \notag \\
&&P(n,k)=0,\quad k=3,5,7,...,\quad \text{or}\quad k>2(n+1).
\end{eqnarray}%
At the accumulation point $\mu _{\infty }=3.5699456...$ the period diverges ($%
n\rightarrow \infty $), and the degree distribution of the
Feigenbaum graph at the onset of chaos becomes a (non truncated)
exponential for even values of the degree,
\begin{eqnarray}
&&P(\infty ,k)=\left( \frac{1}{2}\right) ^{k/2},\quad k=2,4,6,...,
\notag
\label{degree_p_1_critico} \\
&&P(\infty ,k)=0,\quad k=3,5,7,...
\end{eqnarray}%
In figure \ref{distkmucritica} we show the accuracy with which
this analytical result is reproduced by a finite series of
$10^{6}$
data. Numerical and theoretical distributions are in good agreement.%
\newline

\subsubsection{Mean degree $\bar{k}(n)$ and normalized distance $\bar{d}(n)$}

The mean degree $\bar{k}(n)$ is the first moment of the degree
distribution
\begin{equation}
\bar{k}(n)=\sum_{k=2}^{2(n+1)}{kP(n,k)}=4\bigg(1-\frac{1}{2^{n+1}}\bigg),
\end{equation}%
which at the accumulation point yields $\bar{k}(\infty )=4$,
indeed the maximal degree for an HV graph.\newline

Furthermore, by induction it can be shown that the mean distance
$D(n,N)$ within a Feigenbaum graph of $N$ nodes after $n$
period-doubling bifurcations is given approximately by
\begin{equation}
\bar{D}(n,N)\approx \frac{1}{2^{n}}\left\{
\frac{N+1}{3}+\frac{1}{9}\left[ (6n-1)2^{n}+(-1)^{n}\right]
\right\}.  \label{D_n}
\end{equation}%
Observe that for a fixed period $T=2^{n}$ the mean distance
increases linearly with the size $N$ of the graph. Normalization
of this measure in
the limit of infinite size leads to a well-defined mean distance per node $%
\bar d(n)$ with the simple form
\begin{equation}
\bar{d}(n)=\frac{1}{3}\frac{1}{2^{n}},  \label{d_n}
\end{equation}%
such that the linear relation depicted in equation \ref{d_k} is
recovered.

\subsubsection{Clustering coefficient $c(n,k)$}

The local clustering coefficient is a topological measure that
quantify how close a given node's neighbors are to being a clique
\cite{redes,redes2,redes3}. The local clustering coefficient of a
Feigenbaum graph that corresponds to an attractor of period
$T=2^{n}$ is given by
\begin{eqnarray}
&&c(n,k)=\frac{k-1}{{\binom{k}{2}}}=\frac{2}{k},\quad
k=2,4,6,...,2n,  \notag
\label{degree_c_n} \\
&&c(n,k)=\frac{k-2}{{\binom{k}{2}}}=\frac{2(k-2)}{k(k-1)},\quad
k=2(n+1),
\end{eqnarray}%
a result indicative of a so-called hierarchical structure
\cite{hierar}. Since the degree distribution $P(n,k)$ is known in
closed form
throughout the period-doubling cascade, use of the approximation $%
c(n,2(n+1))\approx 1/(n+1)$ leads to
\begin{eqnarray}
&&P(n,k)=\left( {\frac{1}{2}}\right) ^{k/2}=P(n,2/c(n,k)),\quad
k=2,4,6,...,2n,2(n+1),  \notag  \label{degree_p_n_c_n} \\
&&P(n,c)=\left( {\frac{1}{2}}\right) ^{1/c}.
\end{eqnarray}%
Consequently, the mean clustering coefficient $\bar{c}(n)$ is
given by
\begin{eqnarray}
&&\bar{c}(n)=\sum_{k=2}^{2(n+1)}{c(n,k)P(n,k)}=\sum_{k=2}^{2n}{\frac{2}{k}%
\left( {\frac{1}{2}}\right) ^{k/2}}+\frac{2n}{(n+1)(2n+1)}\left( {\frac{1}{2}%
}\right) ^{n}  \notag  \label{c_n_media} \\
&=&\sum_{m=1}^{n}{\frac{1}{m2^{m}}}+\frac{n}{2^{n-1}(n+1)(2n+1)}.
\end{eqnarray}%
This last summation does not possess a solution in closed form
except in the limit $n\rightarrow \infty $ which yields
$\bar{c}(\infty )=\log 2=0.693..$.
Nevertheless, the series converges to $\log 2$ extremely fast: $%
c(0)=0,c(1)=0.666...,c(2)=0.691666...,c(3)=0.693452...$, which
suggests that $\bar{c}(n)$ rapidly loses its dependence on the
bifurcation order $n$ and remains basically constant for all $n$.

\subsubsection{Higher moments of the degree distribution: variance $\sigma ^{2}(n)$}

The moments of the degree distribution $P(n,k)$ can be easily
calculated by making use of the generating function
\begin{equation*}
M(t)=\langle e^{tk}\rangle=1+\left( 4-\frac{2}{2^{n}}\right) t+
\end{equation*}%
\begin{equation*}
\left( 12-\frac{10+4n}{{2}^{n}}\right) {t}^{2}+\left( \frac{104}{3}-\frac{%
(100/3)+20n+4n^{2}}{2^{n}}\right) {t}^{3}+O\left( {t}^{4}\right) .
\end{equation*}%
In particular, the variance is given by
\begin{equation*}
\sigma ^{2}(n)=8-\frac{4+8n}{2^{n}}-{4}^{1-n},
\end{equation*}%
which at the accumulation point becomes $\sigma ^{2}(\infty )=8$.

\section{Reverse bifurcation cascade of chaotic bands: results}

%as the chaotic nature of the attractor is encoded, by construction, in the connectivity pattern associated to the top chaotic band. This %contribution appears coarse-grained in the cumulative distribution $P_{\mu}(n,k\geq 2(n+1))$. The HV algorithm filters out chaotic motion within all %bands except for that taking place in the top band. The contribution to the degree distribution of the chaotic nature of the top band decreases as %$n \to \infty$, and, as it should be expected, at the accumulation point $\mu_{\infty}$ we recover the exponential degree distribution (equation %\ref{degree_p_1_critico}), \textit{i.e.} $\lim_{n\rightarrow\infty}P_{\mu}(n,k)=P(\infty,k)$.\\

\begin{figure}[!htb]
\centering
\includegraphics[width=0.8\textwidth]{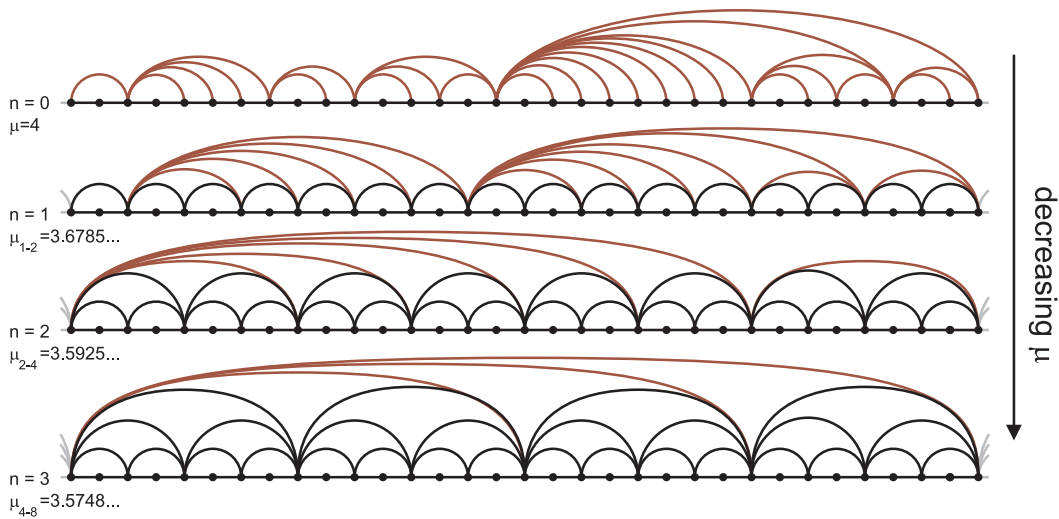}
\caption{Aperiodic Feigenbaum graphs for $\mu>\mu_\infty$. A
sequence of graphs associated with chaotic series after $n$
chaotic-band reverse bifurcations, starting at $\mu=4$ for $n=0$,
when the attractor extends along a single band and the degree
distribution does not present any regularity (red links). For
$n>0$ the phase space is partitioned in $2^n$ disconnected chaotic
bands and the $n$-th self-affine image of $\mu=4$ is the $n$-th
Misiurewicz point $\mu_{2^{n-1}-2^n}$. In all cases, the orbit
visits each chaotic band in the same order as in the periodic
region $\mu<\mu_{\infty}$. This order of visits induces an ordered
structure in the graphs (black links) analogous to that found for
the period-doubling cascade.} \label{grafos de feigenbaum2}
\end{figure}

The Logistic map at $\mu =4$ is said to be fully chaotic as the
Lyapunov exponent attains its maximum value of $\log 2$ there, and
also the single-band attractor spans the unit interval. As $\mu $
is decreased smoothly from $\mu =4$ towards $\mu _{\infty }$, this
first chaotic band ($n=0$) suffers a contraction and a series of
successive band splittings (see figure \ref{figintro}) that are in
several respects analogous to the period-doubling bifurcations at
$\mu <\mu _{\infty }$. Since the band splittings occur from right
to left, they
are called reverse bifurcations \cite{chaos, chaos2, steve, schroeder}. After the first reverse bifurcation ($n=1$%
), the first chaotic band splits into two disconnected chaotic
bands, such that the iteration of an initial condition in this
zone generates an orbit that alternates between these two bands.
While the exact location of each position gives evidence of
sensitive dependence on initial conditions the band alternation is
fixed. Significantly, while in the chaotic zone orbits are
aperiodic, for reasons of continuity they visit each of the $2^n$
chaotic bands in the same order as positions are visited in the
attractors of period $T=2^n$ \cite{schroeder}. In figure
\ref{grafos de feigenbaum2} we have plotted the Feigenbaum graphs
generated through chaotic time series at different values of $\mu$
that correspond to an increasing number of reverse bifurcations.

\begin{figure}[h]
\centering
\includegraphics[width=0.6\textwidth, angle=-90]{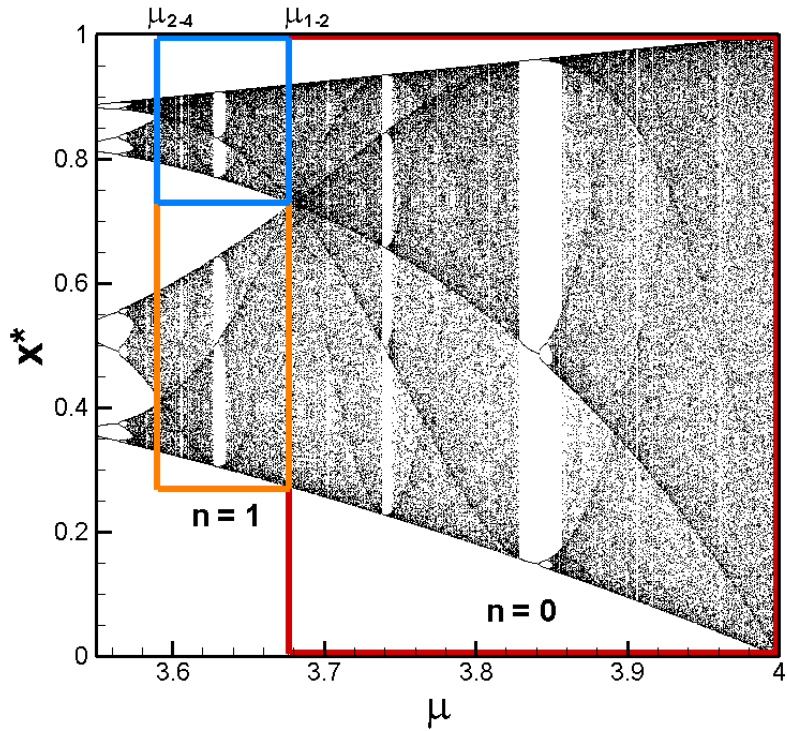}
\caption{{\protect\small {Self-affinity in the chaotic region: the
two
disconnected chaotic bands at $\protect\mu _{2-4}<\protect\mu <\protect\mu %
_{1-2}$} are rescaled copies of the first chaotic band $\protect\mu _{1-2}<%
\protect\mu <4$. An orbit at $\protect\mu _{2-4}<\protect\mu <\protect\mu %
_{1-2}$ makes an alternating journey between both bands.}}
\label{bandas}
\end{figure}

The first of the reverse bifurcation points, called Misiurewicz
points, is located at $\mu
_{1-2}= 3.6785...$. As in the period-doubling region $\mu <\mu _{\infty }$%
, a cascade of chaotic band reverse bifurcations takes place in
the complementary region $\mu >\mu _{\infty }$, such that after
$n$ bifurcations the attractor consists of $2^{n}$ non-overlapping
chaotic bands. The location of the Misiurewicz points converges to
$\mu _{\infty }$ in the limit $n\rightarrow \infty $, and their
relative locations also obey the same asymptotic ratio $\delta
=4.669...$ (Feigenbaum constant for quadratic maps) as $n$
increases as that occurring for the period-doubling bifurcations
at $\mu <\mu _{\infty }$ \cite{chaos, grosman}.\newline
Furthermore, the chaotic bands that are formed after each reverse
bifurcation are self-affine copies of the previous ones
\cite{mandelbrot}. This is illustrated in figure \ref{bandas},
where it can be observed that the two chaotic bands (blue and
orange boxes) appearing after the first reverse bifurcation (at
$\mu _{1-2}$) are rescaled copies of the first chaotic band (red
box). Accordingly, each value of $\mu $ within the first chaotic
band has $n$ self-affine images after $n$ reverse bifurcations.
For instance, the
first Misiurewicz point $\mu _{1-2}=3.6785...$ is a self-affine image of $%
\mu =4$ after the first reverse bifurcation, the second Misiurewicz point $%
\mu _{2-4}=3.5925...$ is also a self-affine image of $\mu =4$
after the second reverse bifurcation, and so on. Consider now a
chaotic time series extracted at $\mu \in \lbrack \mu _{2-4}-\mu
_{1-2}]$ in which the iterate position values alternate between
both chaotic bands. Now separate this series into two time-ordered
subseries, each of which contain only data belonging to either the
top or the bottom chaotic bands. These subseries are indeed
rescaled copies of a series extracted from the first chaotic band
at the corresponding preimage value of $\mu $. Notice that the HV
algorithm is invariant under affine transformations in the series.
Accordingly, the Feigenbaum graphs of these subseries are the same
as the graph obtained for the self-affine value of $\mu $
belonging to the first chaotic band.

\subsection{Self-affine properties of chaotic bands: mean degree and degree distribution}

As mentioned, the interband motion that an orbit experiences
within an attractor composed of $2^{n}$ chaotic bands follows the
same order of visits as that we described for a periodic orbit of
period $2^{n}$. And as in the case of the periodic region, this
property possibilitates the derivation of the main characteristics
of the Feigenbaum graphs in the chaotic region. Consider the
Feigenbaum graph of a chaotic series generated by the Logistic map
after $n$ reverse bifurcations. By construction, the structure of
this graph is in many respects a mirror image of the Feigenbaum
graph in the periodic region after $n$ period-doubling
bifurcations. The only differences originate from the data located
within the top chaotic band. This feature is due to the fact that
chaotic bands do not overlap and that the HV algorithm filters out
the precise locations of data in favor of unspecified relative
positions. Denote by $P_{\mu }(n,k)$ its degree distribution,
where, in order to not overburden
the notation, we refer to $\mu$ indistinctly of the self-affine image of $%
\mu $ we study. By construction, and up to $k=2n$, the degree
distribution $P_{\mu }(n,k)$ is equivalent to its periodic region
counterpart, i.e.
\begin{equation}
P_{\mu }(n,k)=\left( \frac{1}{2}\right) ^{k/2},\quad
k=2,4,6,...,2n. \label{degree_p_mu_supp}
\end{equation}%
Now, the data belonging to the top chaotic band transmits a
specific contribution to the associated Feigenbaum graph which in
general cannot be determined analytically, and as we shall see
shows
up in $P_{\mu }(n,k)$ for $k>2n$. This contribution is denoted $P_{\mu }^{%
\text{top}}(n,k)$, where, by definition, $P_{\mu
}^{\text{top}}(0,k)=P_{\mu }(0,k)$, the degree distribution of the
Feigenbaum graph at the first
chaotic band $n=0$, $\mu _{1-2}<\mu \leq 4$. While the precise shape of the $%
\mu $-dependent distribution $P_{\mu }^{\text{top}}(n,k)$ is
unknown, the self-affine structure of the band-splitting cascade
described previously makes it possible to relate it to the first
chaotic band, i.e.
\begin{equation*}
P_{\mu }^{\text{top}}(n,k)=P_{\mu }(0,k).
\end{equation*}%
This important property, that, as we shall comment below,
corresponds to a crossover phenomenon in the RG flows, breaks down
the structure of the Feigenbaum graphs within the chaotic region
into two contributions: (i) an interband contribution stemming
from the order of visits amongst chaotic bands that is equivalent
to that for periodic attractors (with associated black links in
figure \ref{grafos de feigenbaum2}), and (ii) a contribution from
the first chaotic band (with associated red links in figure
\ref{grafos de feigenbaum2}). For normalization reasons, this
second contribution can be coarse-grained as a cumulative
distribution, leading to
\begin{eqnarray}
&&P_{\mu }(n,k)=\left( \frac{1}{2}\right) ^{k/2},\quad
k=2,4,6,...,2n,
\notag  \label{degree_p_mu_supp2} \\
&&P_{\mu }(n,k\geq 2(n+1))=\left( \frac{1}{2}\right) ^{n}.
\end{eqnarray}%
Interestingly, as $n$ increases the contributions from the
interband motion of the chaotic attractor these become more and
more dominant and the contribution associated with the first
chaotic band is progressively smeared out, disappearing at the
accumulation point $\mu _{\infty }$ (see figure \ref{grafos de
feigenbaum2}).

Based on the previous arguments on self-affinity, the contribution
from the first chaotic band can be determined quantitatively as
follows. For concreteness, let us focus on a chaotic attractor
located after the first reverse bifurcation. The attractor is
divided into two non-overlapping chaotic bands, and every orbit in
it alternates between the `top' and `bottom' bands. While the
precise location of the visits within each band is unknown, by
construction the nodes associated to the data located in the
`bottom' band have degree $k=2$, and this occurs for half of the
iteration times, so
\begin{equation*}
P_{\mu }(1,2)=\frac{1}{2}.
\end{equation*}%
Also, for parity reasons (even number of chaotic bands) one has
\begin{equation*}
P_{\mu }(1,3)=0,
\end{equation*}%
independently of $\mu $. For $k>3$, the contribution comes
necessarily from the top band, and normalization yields
\begin{equation*}
P_{\mu }(1,k)=\frac{1}{2}P_{\mu }^{\text{top}}(1,k-2),\quad k\geq
4.
\end{equation*}%
Finally, self-affinity implies
\begin{equation*}
P_{\mu }^{\text{top}}(1,k)=P_{\mu }(0,k).
\end{equation*}%
Repeating this same procedure, we find that after $n$ reverse
bifurcations the degree distribution of a Feigenbaum graph
fulfills the expressions
\begin{equation}
P_{\mu }(n,k)=\left( \frac{1}{2}\right) ^{k/2},\quad
k=2,4,6,...,2n, \label{degdist1}
\end{equation}%
\begin{equation}
P_{\mu }(n,k)=0,\quad k=3,5,7,...,2n+1,  \label{degdist2}
\end{equation}%
independently of $\mu $, and
\begin{equation}
P_{\mu }(n,k)=\left( \frac{1}{2}\right) ^{n}P_{\mu }^{\text{top}%
}(n,k-2n)=\left( \frac{1}{2}\right) ^{n}P_{\mu }(0,k-2n),\quad
k\geq 2(n+1), \label{degdist3}
\end{equation}%
that depends on $\mu $. As expected, the last expression only
depends on the structure of the first chaotic band, decreases for
increasing values of $n$ and disappears in the limit $n\rightarrow
\infty $.\\

\noindent The above expressions for $P_{\mu }(n,k)$ can be used to
derive the mean degree $\bar{k}_{\mu }(n)$ of a Feigenbaum graph
in the chaotic regime:
\begin{eqnarray}
\bar{k}_{\mu }(n)=\sum_{k=2}^{2n}{k\left( {\frac{1}{2}}\right) ^{k/2}}%
+\left( \frac{1}{2}\right) ^{n}\sum_{k=2(n+1)}^{\infty }{\ kP_{\mu
}(0,k-2n)}=4\left( 1-\frac{1}{2^{n}}\right) +\frac{\bar{k}_{\mu
}(0)}{2^{n}}. \label{kmedia1}
\end{eqnarray}%
This last expression relates the mean degree after $n$ reverse
bifurcations to that in the first chaotic band. Interestingly, the
only solution with constant mean degree in the chaotic regime is
$\bar{k}_\mu(0)=\bar{k}_\mu(n)=4$, in full agreement with the
general theory.\newline

\subsection{Periodic windows: self-affine copies of the Feingenbaum
diagram}

\begin{figure}[h]
\centering
\includegraphics[width=0.95\textwidth]{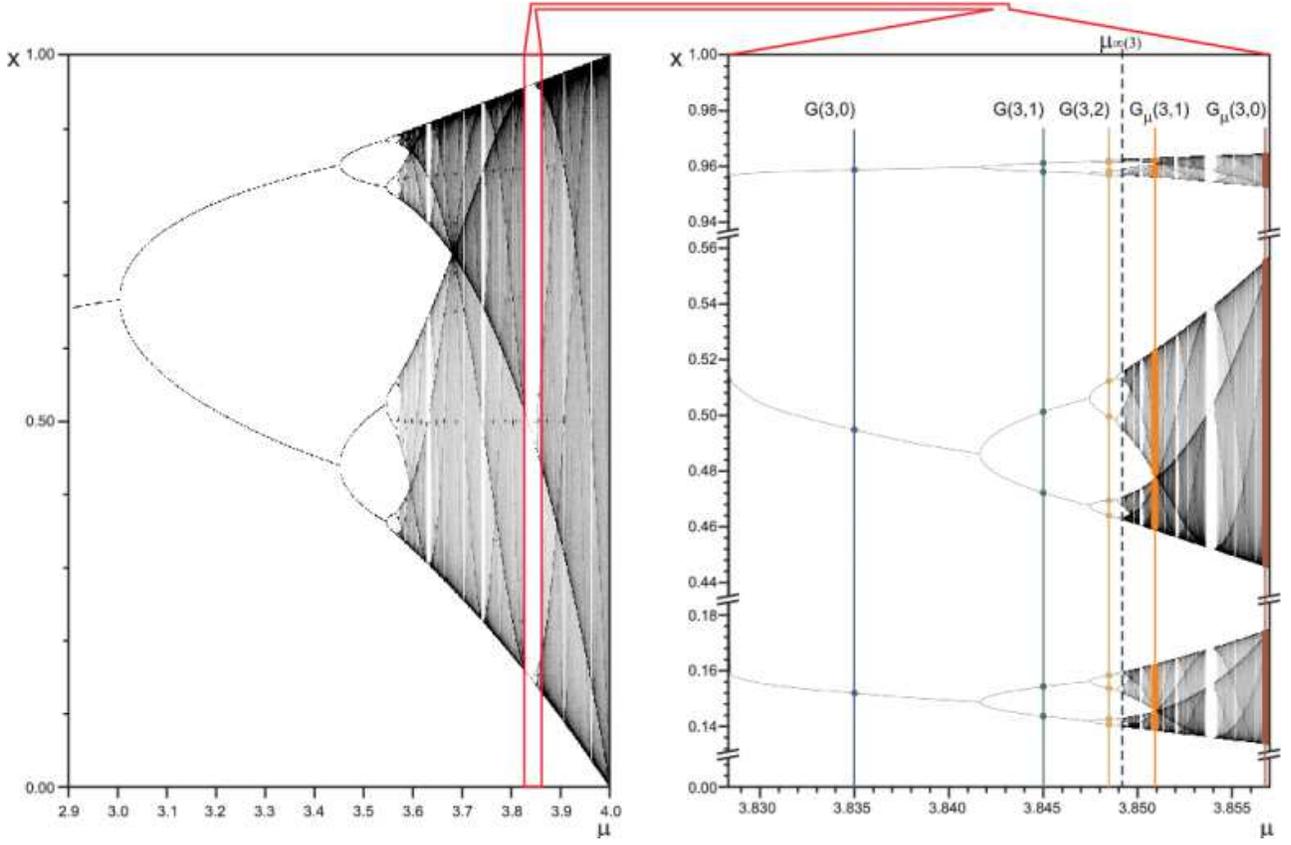}
\caption{Zoom of the Feigenbaum diagram close to the period $m=3$
window. Starting with a period $3$ orbit, each one of the stable
branches develops into a period-doubling bifurcation cascade with
a new accumulation point $\mu_\infty(3)$, beyond which the
attractor becomes chaotic, interwoven with periodic windows: each
part of the diagram is indeed a rescaled copy of the full
Feigenbaum tree. The locations of several Feigenbaum graphs (with
the notation defined in the text) within the period three window
are depicted.} \label{windows}
\end{figure}
The Feigenbaum diagram shows a rich self-affine structure:
periodic windows of initial period $m$ that undergo successive
period-doubling bifurcations with new accumulation points
$\mu_{\infty}(m)$ appear interwoven with chaotic attractors at
$\mu>\mu_{\infty}$. These period-doubling cascades taking place
beyond $\mu_{\infty}$ are self-affine copies of the fundamental
period-doubling route to chaos ($\mu<\mu_{\infty}$). For instance,
the window that initiates with period $m=3$ ($\mu\approx3.84$)
generates three period-doubling cascades, each one being a
properly rescaled copy of the fundamental one (see figure
\ref{windows} for a graphical illustration). In order to take into
account such phenomenology in the labeling of Feigenbaum graphs,
we make use of the following notation: $G(m,n)$ is the Feigenbaum
graph associated with a periodic attractor of period
$T=m\cdot2^n$, that is, the graph belonging to a periodic window
of initial period $m$, after $n$ period-doubling bifurcations.
Observe that the process of chaotic band reverse bifurcations that
take place from $\mu=4$ towards $\mu_{\infty}$ is again
 repeated in a self-affine manner, indeed, each accumulation point $\mu_{\infty}(m)$ is the limiting value
 of a chaotic-band reverse bifurcation cascade. In figure \ref{windows} this aspect is also illustrated.
 Accordingly, we may also extend the notation of Feigenbaum graphs belonging to chaotic regions,
 such that $G_{\mu}(m,n)$ is associated with a chaotic attractor composed by $m \cdot 2^n$ bands
 (that is, after $n$ reverse bifurcations of $m$ initial chaotic
 bands).\\

\noindent Given a periodic window with initial period $m$ within
the first chaotic band ($\mu _{1-2}<\mu \leq 4$), we note that
after $n$ reverse bifurcations, the chaotic-band attractors
contain a self-affine image of that window, where there is an
initial periodic attractor of period $m\cdot 2^{n}$ developing
into $m\cdot 2^{n}$ period-doubling cascades inside the window. In
this case, equation \ref{kmedia1} reads
\begin{equation*}
\bar{k}_{\mu }(n,m\cdot 2^{n})=4\left( 1-\frac{1}{2^{n}}\right) +\frac{\bar{k%
}_{\mu }(0,m)}{2^{n}},
\end{equation*}%
that together with equation \ref{th1} yields
\begin{equation*}
\bar{k}_{\mu }(n,m\cdot 2^{n})=4\bigg(1-\frac{1}{2(m2^{n})}\bigg),
\end{equation*}%
as expected.

\section{Renormalization Group approach}

\begin{figure}[!htb]
\centering
\includegraphics[width=0.9\textwidth]{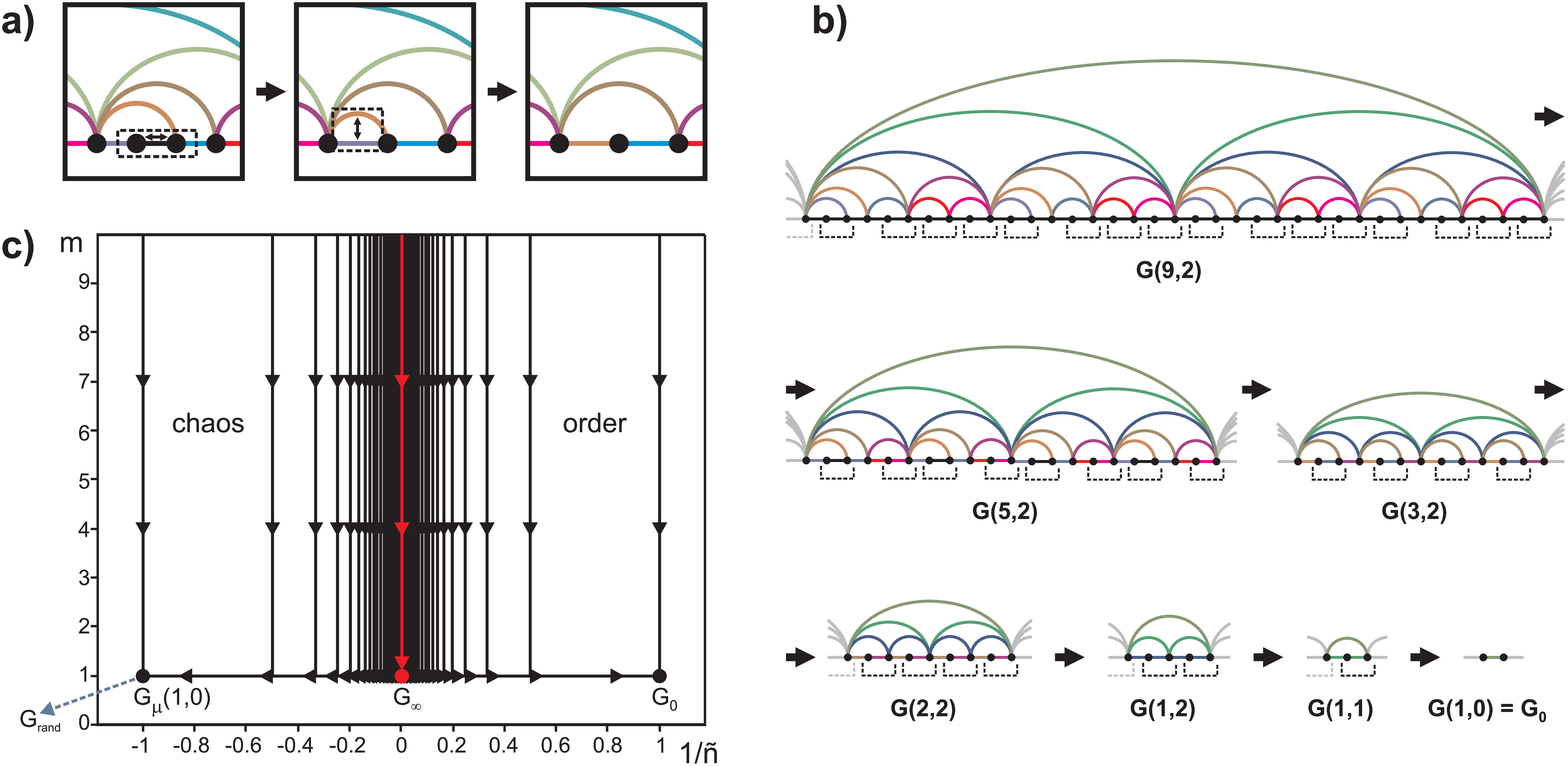}
\caption{Renormalization process and network RG flow structure.
(a) Illustration of the renormalization process $\cal R$: a node
with degree $k=2$ is coarse-grained with one of its neighbors
(indistinctively) into a block node that inherits the links of
both nodes. This process coarse-grains every node with degree
$k=2$ present at each renormalization step. (b) Example of an
iterated renormalization process in a sample Feigenbaum graph at a
periodic window with initial period $m=9$ after $n=2$
period-doubling bifurcations (an orbit of period
$T=m\cdot2^n=36$). (c) RG flow diagram, where $m$ identifies the
periodic window that is initiated with period $m$ and
\textit{\~{n}} designates the order of the bifurcation,
\textit{\~{n}} $=n+1$ for period-doubling bifurcations and
\textit{\~{n}} $=-(n+1)$ for reverse bifurcations. $\Delta
\mu(m)=\mu_\infty(m)-\mu$ denotes the reduced control parameter of
the map, and $\mu_\infty(m)$ is the location of the accumulation
point of the bifurcation cascades within that window. Feigenbaum
graphs associated with periodic series ($\Delta \mu (m)>0$,
\textit{\~{n}} $>0$) converge to $G(1,0)\equiv G_0$ under the RG,
whereas those associated with aperiodic ones ($\Delta\mu(m)<0$,
\textit{\~{n}} $<0$) converge to $G_{\text{rand}}$. The
accumulation point $\mu_\infty\equiv\mu_\infty(1)$ corresponds to
the unstable (nontrivial) fixed point $G(1,\infty)\equiv G_\infty$
of the RG flow, which is nonetheless approached through the
critical manifold of graphs $G(m,\infty)$ at the accumulation
points $\mu_\infty(m)$. In summary, the nontrivial fixed point of
the RG flow is only reached via the family of the accumulation
points, otherwise the flow converges to trivial fixed points for
periodic or chaotic regions.} \label{RGfig}
\end{figure}
\subsection{RG transformation: definition, flows and fixed points}
 In order to recast previous findings in the context of the renormalization group, let us define an RG operation $\cal R$ on a graph as the
 coarse-graining of every couple of adjacent nodes where one of them has degree $k=2$ into a block node that inherits the links of the previous
 two nodes (see figure \ref{RGfig}.a). This is a real-space RG transformation on the Feigenbaum graph \cite{newmannRG}, dissimilar from recently
 suggested box-covering complex network renormalization schemes \cite{rg, rg2, rg3}. As a matter of fact, this scheme turns out to be equivalent for
 $\mu<\mu_{\infty}$ to the construction of an HV graph from the composed map $f^{(2)}$ instead of the original $f$, in correspondence
 to the original Feigenbaum renormalization procedure \cite{Feigenbaum, steve}. We first note that ${\cal R}\{G(1,n)\}=G(1,n-1)$, thus,
 an iteration of this process yields an RG flow that converges to the (1st) trivial fixed point
 ${\cal R}^{(n)}\{G(1,n)\}= G(1,0)\equiv G_0= {\cal R}\{G_0 \}$. This is the stable fixed point of the RG
 flow $\forall \mu<\mu_{\infty}$. We note that there is
  only one relevant variable in our RG scheme, represented by the reduced control parameter $\Delta \mu=\mu_{\infty}-\mu$, hence, to identify a
  nontrivial fixed point we set $\Delta \mu=0$ or equivalently $n \to \infty$, where the structure of the Feigenbaum graph turns to be
  completely self-similar under $\cal R$. Therefore we conclude that $G(1,\infty)\equiv G_\infty$ is the nontrivial fixed point of the
  RG flow, ${\cal R}\{G_\infty \}=G_\infty$. In connection with this, let $P_t(k)$ be the degree distribution of a generic Feigenbaum
  graph $G_t$ in the period-doubling cascade after $t$ iterations of $\cal R$, and point out that the RG operation, ${\cal R}\{G_t\}=G_{t+1}$,
  implies a recurrence relation $(1-P_t(2))P_{t+1}(k)=P_t(k+2)$, whose fixed point coincides with the degree distribution found in
  equation \ref{degree_p_1_critico}. This confirms that the nontrivial fixed point of the flow is indeed $G_\infty$.\\

\noindent Next, under the same RG transformation, the self-affine
structure of the family of chaotic attractors yields ${\cal
R}\{G_{\mu}(1,n)\}=G_{\mu}(1,n-1)$, generating a RG flow that
converges to the Feigenbaum graph associated to the 1st chaotic
band, ${\cal R}^{(n)}\{G_{\mu}(1,n) \}=G_{\mu}(1,0)$. Repeated
application of $\cal R$ breaks temporal correlations in the
series, and the RG flow leads to a 2nd trivial fixed point ${\cal
R}^{(\infty)}\{G_{\mu}(1,0)\}=G_{\text{rand}}={\cal
R}\{G_{\text{rand}}\}$, where $G_{\text{rand}}$ is the HV graph
generated by a purely uncorrelated random process. This graph has
a universal degree distribution $P(k)=(1/3)(2/3)^{k-2}$,
independent of the random process
underlying probability density (see \cite{pre, submitted}).\\

\noindent Finally, let us consider the RG flow inside a given
periodic window of initial period $m$. As the renormalization
process addresses nodes with degree $k=2$, the initial
applications of $\cal R$ only change the core structure of the
graph associated with the specific value $m$ (see figure
\ref{RGfig}.b for an illustrative example). The RG flow will
therefore converge to the 1st trivial fixed point via the initial
path ${\cal R}^{(p)}\{G(m,n)\}=G(1,n)$, with $p \le m$, whereas it
converges to the 2nd trivial fixed point for $G_{\mu}(m,n)$ via
${\cal R}^{(p)}\{G_{\mu}(m,n)\}=G_{\mu}(1,n)$. In the limit of
$n\rightarrow\infty$ the RG flow proceeds towards the nontrivial
fixed point via the path ${\cal
R}^{(p)}\{G(m,\infty)\}=G(1,\infty)$. Incidentally, extending the
definition of the reduced control parameter to $\Delta
\mu(m)=\mu_\infty(m)-\mu$, the family of accumulation points is
found at $\Delta\mu(m)=0$. A complete
schematic representation of the RG flows can be seen in figure \ref{RGfig}.c.\\

\noindent Interestingly, and at odds with standard RG applications
to (asymptotically) scale-invariant systems, we find that
invariance at $\Delta \mu=0$ is associated in this instance to an
exponential (rather than power-law) function of the observables,
concretely, that for the degree distribution. The reason is
straightforward: ${\cal R}$ is not a conformal transformation
($\textit{i.e.}$ a scale operation) as in the typical RG, but
rather, a translation procedure. The associated invariant
functions are therefore non homogeneous (with
the property $\text{g}(ax)=b\text{g}(x)$), but exponential (with the property $\text{g}(x+a)=c\text{g}(x)$).\\

\subsection{Crossover phenomenon}

\begin{figure}[h]
\centering
\includegraphics[width=0.9\textwidth]{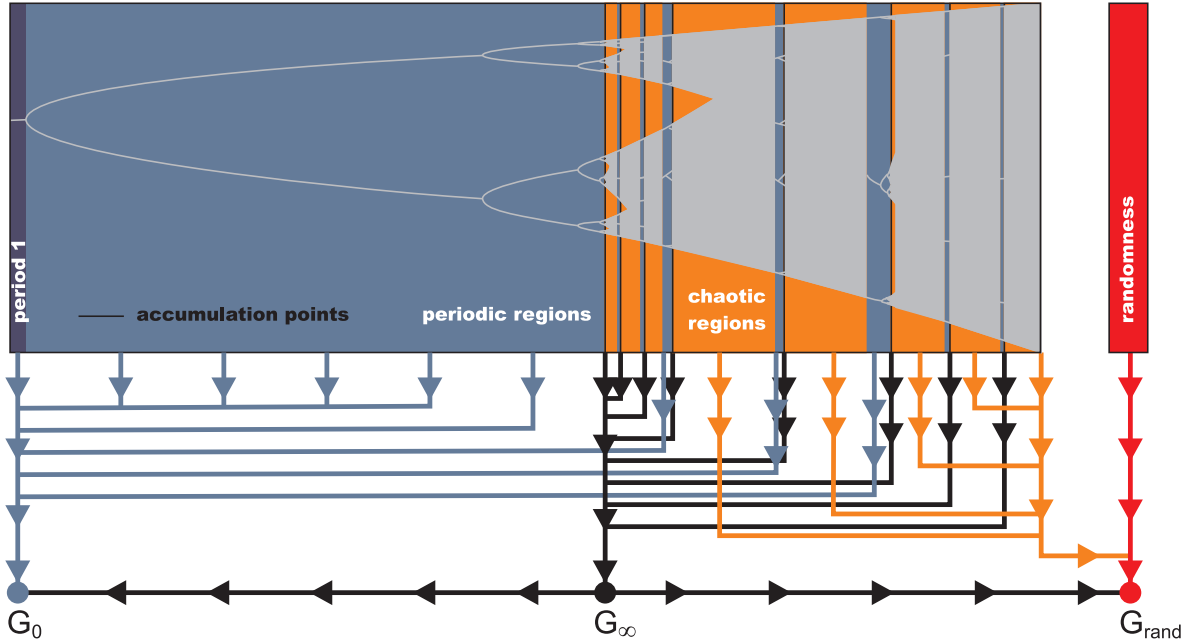}
\caption{Illustrative cartoon incorporating the RG flow of
Feigenbaum graphs along the entire Feigenbaum diagram: aperiodic
(chaotic or random) series generate graphs whose RG flow converges
to the trivial fixed point $G_{\text{rand}}$, whereas periodic
series (both in the region $\mu<\mu_{\infty}$ and inside periodic
windows) generate graphs whose RG flow converges to the trivial
fixed point $G(0,1)$. The nontrivial fixed point of the RG flow
$G(\infty,1)$ is only reached through the critical manifold of
graphs at the accumulation points $\mu_{\infty}(m)$.}
\label{warhol}
\end{figure}

When the RG transformation for a Feigenbaum graph is applied
repeatedly, it generates flows terminating at two different
trivial fixed points $G(1,0)\equiv G_0$ and $G_{\text{rand}}$ or
at a nontrivial fixed point $G(1,\infty)\equiv G_\infty$. $G_0$ is
a chain graph where every node has two links, $G_{\text{rand}}$ is
a graph associated with a purely random uncorrelated process,
whereas $G_\infty$ is a self-similar graph that represents the
onset of chaos. The RG properties within the periodic windows are
incorporated into a general RG flow diagram (see figure
\ref{RGfig}.c and figure \ref{warhol} for an alternative
representation). Here we add a comment on the standard presence of
crossover phenomena in the RG applications, in our case for large $n$ (or $%
\mu \simeq \mu _{\infty }$) for both $\mu <\mu _{\infty }$ and
$\mu >\mu
_{\infty }$. In both cases the graphs $G(1,n-j)$ and $G_{\mu }(1,n-j)$ with $%
j\ll n$ closely resemble the self-similar $G_\infty$ (obtained only when $%
\mu =\mu _{\infty }$) for a range of values of the number $j$ of
repeated applications of the transformation $\cal R$ until a clear
departure takes place towards $G_0$ or $G_{\text{rand}}$ when $j$
becomes comparable to $n$. Hence, for instance the graph ${\cal
R}^{(j)}\{G_\mu(1,n)\}$ will only show its true chaotic nature
(and therefore converge to $G_{\text{rand}}$) once $j$ and $n$ are
of the same order. In other words, this happens once its degree
distribution becomes dominated by the contribution of $P_{\mu
}^{\text{top}}(n,k)$ (alternatively, once the graph core -related
to the chaotic band structure and the order of visits to chaotic
bands- is removed by the iteration of the renormalization
process).

\section{Network entropy}
\subsection{Entropy optimization and RG fixed points}
Recently \cite{robledo}, it has been point out that there exists a
connection between the extremal properties of entropy expressions
and the renormalization group approach, when applied to systems
with scaling symmetry. Namely, that the fixed points of RG flows
can be obtained through a process of entropy optimization,
providing the RG approach a variational flavor. In this section we
investigate on this connection, and derive, via optimization of an
entropic functional for the Feigenbaum graphs, all the RG flow
directions and fixed points directly from the information
contained in the degree distribution. We depart by applying
Jayne's principle of Maximum Entropy \cite{jaynes} with two
different restrictions and prove that the degree distribution
$P(k)$ that maximizes its entropy $h$ is in each case related to
the fixed points of the RG flow. We employ the standard
optimization technique that involves
Lagrange multipliers.\\

\noindent Let us begin by defining a graph entropy as the Shannon
entropy of the graph degree distribution
\begin{equation}
h=-\sum_{k=2}^{\infty }{P(k)\log {P(k)}},  \label{entropia}
\end{equation}%
and assume that the degree distribution ${P(k)}$ has a well-defined mean $%
\bar{k}$ (this is true in general for HV graphs and in particular
for Feigenbaum graphs according to equation \ref{th1}). The,
consider the Lagrangian
\begin{equation*}
\mathcal{L}=-\sum_{k=2}^{\infty }{P(k)\log {P(k)}}-(\lambda
_{0}-1)\left(
\sum_{k=2}^{\infty }{P(k)}-1\right) -\lambda _{1}\left( \sum_{k=2}^{\infty }{%
kP(k)}-\bar{k}\right) ,
\end{equation*}%
for which the extremum condition reads
\begin{equation*}
\frac{\partial \mathcal{L}}{\partial P(k)}=-\log {P(k)}-\lambda
_{0}-\lambda _{1}k=0,
\end{equation*}%
and has the general solution
\begin{equation*}
P(k)=e^{-\lambda _{0}-\lambda _{1}k}.
\end{equation*}%
The Lagrange multipliers $\lambda _{0}$ and $\lambda _{1}$ can be
calculated from their associated constraints. First, the
normalization of the probability density,
\begin{equation*}
\sum_{k=2}^{\infty }{e^{-\lambda _{0}-\lambda _{1}k}}=1,
\end{equation*}%
implies the following relation between $\lambda _{0}$ and
$\lambda_{1}$
\begin{equation*}
e^{\lambda _{0}}=\sum_{k=2}^{\infty }{e^{-\lambda
_{1}k}}=\frac{e^{-\lambda _{1}}}{e^{\lambda _{1}}-1},
\end{equation*}%
and differentiation of this last expression with respect to
$\lambda_{1}$ yields
\begin{equation*}
-\sum_{k=2}^{\infty }{ke^{-\lambda _{1}k}}=\frac{e^{-\lambda _{1}}-2}{%
(e^{\lambda _{1}}-1)^{2}}.
\end{equation*}%
Second, the assumption that the mean degree is a well-defined
quantity (true for HV graphs) yields
\begin{equation*}
\sum_{k=2}^{\infty }{ke^{-\lambda _{0}-\lambda _{1}k}}=\bar{k}=\frac{%
2-e^{-\lambda _{1}}}{1-e^{-\lambda _{1}}}.
\end{equation*}%
Combining the above results we find
\begin{equation*}
\lambda _{1}=\log \left( \frac{\bar{k}-1}{\bar{k}-2}\right) ,
\end{equation*}%
and%
\begin{equation*}
\lambda _{0}=\log \left( \frac{(\bar{k}-2)^{2}}{\bar{k}-1}\right)
.
\end{equation*}%
Hence, the degree distribution that maximizes $h$ is
\begin{equation*}
P(k)=\frac{\bar{k}-1}{(\bar{k}-2)^{2}}\left( \frac{\bar{k}-2}{\bar{k}-1}%
\right) ^{k},
\end{equation*}%
which is an increasing function of $\bar{k}$. The maximal entropy
is therefore found for the maximal mean degree $\bar{k}=4$, with
an associated degree distribution
\begin{equation*}
P(k)=\frac{3}{4}\left( \frac{2}{3}\right) ^{k}=\frac{1}{3}\left( \frac{2}{3}%
\right) ^{k-2}.
\end{equation*}%
Remarkably, we conclude that the HV graph with maximal entropy is
that
associated with a purely uncorrelated random process, and coincides with $G_{%
\text{rand}}$, a trivial fixed point of the RG flow.\newline

Observe now that, by construction, the Feigenbaum graphs along the
period-doubling route to chaos ($\mu<\mu_\infty$) do not have odd
values for the degree. Let us assume this additional constraint in
the former entropy optimization procedure. The derivation proceeds
along similar steps, although summations now run only over even
terms. Concretely, we have
\begin{equation*}
e^{\lambda _{0}}=\sum_{k=1}^{\infty }{e^{-\lambda _{1}2k}}=\frac{1}{%
e^{2\lambda _{1}}-1},
\end{equation*}%
which after differentiation over $\lambda _{1}$ gives
\begin{equation*}
\sum_{k=1}^{\infty }{ke^{-\lambda _{1}2k}}=\frac{e^{2\lambda _{1}}-2}{%
(e^{2\lambda _{1}}-1)^{2}}
\end{equation*}%
and
\begin{equation*}
\sum_{k=1}^{\infty }{2ke^{-\lambda _{0}-\lambda _{1}2k}}=\bar{k}=\frac{%
2e^{2\lambda _{1}}}{e^{2\lambda _{1}}-1}.
\end{equation*}%
We obtain for the Lagrange multipliers
\begin{equation*}
\lambda _{1}=\frac{1}{2}\log \left(
\frac{\bar{k}}{\bar{k}-2}\right)
\end{equation*}%
and
\begin{equation*}
\lambda _{0}=\log \left( \frac{\bar{k}-2}{2}\right) .
\end{equation*}%
The degree distribution that maximizes the graph entropy turns now
to be
\begin{equation*}
P(k)=\frac{2}{\bar{k}-2}\left( \frac{\bar{k}-2}{\bar{k}}\right)
^{k/2}.
\end{equation*}%
As before, entropy is an increasing function of $\bar{k}$,
attaining its
larger value for the upper bound value $\bar{k}=4$, which reduces to $%
P(k)=\left( 1/2\right) ^{k/2}$, $k=2,4,6,...$, equation
\ref{degree_p_1_critico}.
 We conclude that the maximum
entropy of the entire family of Feigenbaum graphs, if we require
that odd values for the degree are not allowed, is achieved at the
accumulation point, that is, the nontrivial fixed point of the RG
flow.\\
Finally, the network entropy is trivially minimized for a degree
distribution $P(2)=1$, that is, at the trivial fixed point $G_0$.
These results indicate that the fixed point structure of a RG flow
can be obtained from an entropy optimization process, confirming
the aforementioned connection.\newline

\subsection{Network entropy and Pesin theorem}

\subsubsection{Periodic attractors} By making use of the expression
we have for the degree distribution $P(n,k)$ in the region $\mu
<\mu _{\infty }$ we obtain for the graph entropy $h(n)$, after the
$n$-th period-doubling bifurcation, the following result
\begin{eqnarray}
&&h(n)=-\sum_{k=2}^{2(n+1)}{P(n,k)\log P(n,k)}  \notag
\label{entropia_m1}
\\
&=&-\sum_{k=2}^{2n}{\frac{1}{2^{k/2}}\log \left( \frac{1}{2^{k/2}}\right) }-%
\frac{1}{2^{n}}\log \left( \frac{1}{2^{n}}\right) =\frac{\log 2}{2}(\bar{k}-%
\frac{2}{2^{n}})=\log 4\left( 1-\frac{1}{2^{n}}\right) .
\end{eqnarray}%
We observe that the entropy increases with $n$ and, interestingly,
depends linearly on the mean degree $\bar{k}$. This linear
dependence between $h$ and $\bar{k}$ is related to the fact that,
generally speaking, the entropy and the mean of a probability
distribution are proportional for exponentially distributed
functions, a property that holds exactly in the accumulation point
(eq.\ref{degree_p_1_critico}) and approximately in the periodic
region (eq.\ref{degree_p_n}), where there is a finite cut off that
ends the exponential law. Interestingly, HV graphs associated with
chaotic series have also been found to have an asymptotically
exponential degree distribution \cite{submitted}. Finally, note
that in the limit $n\rightarrow \infty $ (accumulation point) the
entropy converges to a finite value $h(\infty )=\log 4$.\newline

%Similar arguments can be employed to calculate the graph entropy inside the windows of stability. For instance, in the window of stability with initial %period $m=3$, we have
%\begin{eqnarray}\label{entropia_m1}
%&&h(n)=-\sum_{k=2}^{2n+5}{P(n,k)\log(P(n,k))}\nonumber\\
%&&=-\frac{1}{3}\log\left(\frac{1}{3}\right)-\sum_{k=3}^{2n+3} \frac{1}{3\cdot 2^{\frac{k-3}{2}}}\log\left(\frac{1}{3\cdot 2^{\frac{k-3}%{2}}}\right)-\frac{1}{3\cdot 2^{n}}\log\left(\frac{1}{3 \cdot 2^{n}}\right)\nonumber\\
%&&=\log 4\left(1-\frac{1}{3\cdot 2^n}\right),
%\end{eqnarray}
%this suggests a priori a general expression for the entropy in any period-doubling bifurcation cascade
%$$h(m,n)=\log 4\left(1-\frac{1}{m\cdot 2^n}\right).$$
%This expression is, nevertheless, not true in general for any window of stability with initial period $m$. However, decomposing the graph entropy into %two contributions, namely, (i)  the contribution $\bar h(m)$ of the initial graph motif of period $m$ and (ii) the contribution of a period-%doubling process, the following formula can be written down
%$$h(m,n)=\bar h(m) -\sum_{l=1}^{n}{\frac{1}{m 2^l}\log\left(\frac{1}{m 2^l}\right)}-\frac{1}{m 2^n}\log\left(\frac{1}{m 2^n}\right)$$
%$$h(m,n)=\bar h(m) + \frac{\log(4m)}{m}-\frac{\log 4}{m 2^n}$$
%For each accumulation point we have
%$$h(m,\infty)=\bar h(m) + \frac{\log(4m)}{m}.$$

\subsubsection{Chaotic attractors} For Feigenbaum graphs
$G_\mu(1,n)$ (in the chaotic region), in general $h$ cannot be
derived exactly since the precise shape of $P(k)$ is unknown
(albeit the asymptotic shape is also exponential
\cite{submitted}). However, arguments of self-affinity similar to
those used for describing the degree distribution of Feigenbaum
graphs can be employed in order to find some regularity properties
of the entropy $h_{\mu }(n)$. Concretely, the entropy after $n$
chaotic band reverse bifurcations can be expressed as a function
of $n$ and of the entropy in the first chaotic band $h_{\mu }(0)$.
Using the expression for the degree distribution, a little algebra
yields

\begin{equation*}
h_{\mu }(n)=\log 4+\frac{h_{\mu }^{\text{top}}(n)}{2^{n}}=\log 4+\frac{%
h_{\mu }(0)}{2^{n}}.
\end{equation*}
\begin{figure}[!htb]
\centering
\includegraphics[width=0.7\textwidth]{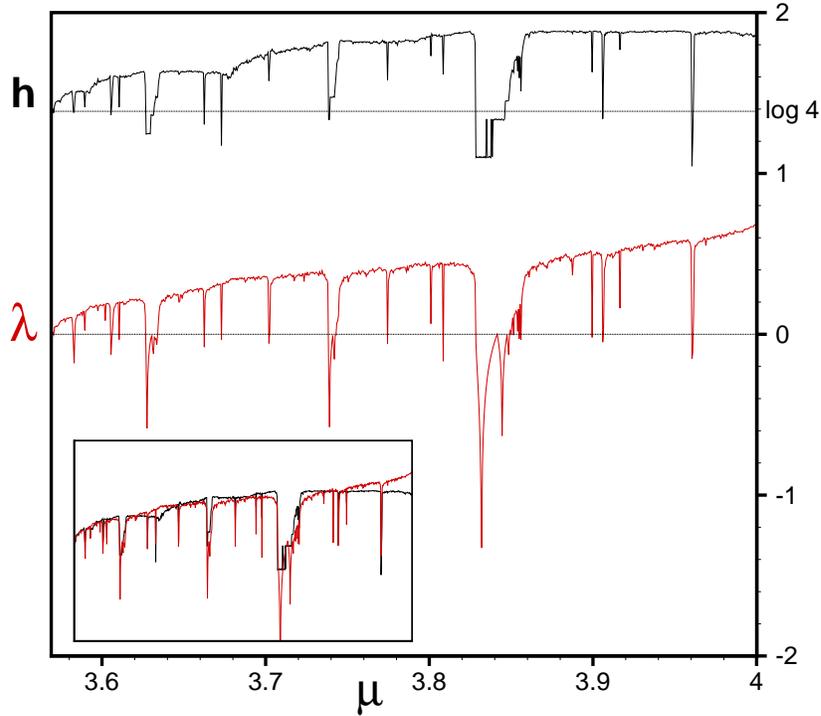}
\caption{Horizontal visibility network entropy $h$ and Lyapunov
exponent $\lambda$ for the Logistic map. We plot the numerical
values of $h$ and $\lambda$ for $3.5<\mu<4$ (the numerical step is
$\delta\mu=5\cdot10^{-4}$ and in each case the processed time
series have a size of $2^{12}$ data). The inset reproduces the
same data but with a rescaled entropy $h-\log(4)$. The
surprisingly good match between both quantities is reminiscent of
the Pesin identity (see text). Unexpectedly, the Lyapunov exponent
within the periodic windows ($\lambda<0$ inside the chaotic
region) is also well captured by $h$.} \label{entropy}
\end{figure}

In figure \ref{figintro} we observe how the chaotic-band reverse
bifurcation process takes place in the chaotic region from right
to left, and therefore leads in this case to a decrease of entropy
with an asymptotic value of $\log 4$ for $n\rightarrow\infty$ at
the accumulation point. These results suggest that the graph
entropy behaves qualitatively as the map's Lyapunov exponent
$\lambda$, with the peculiarity of having a shift of  $\log 4$, as
confirmed numerically in figure \ref{entropy}.
 This unexpected qualitative agreement is reasonable in the chaotic region in view of the
 Pesin theorem \cite{chaos2}, that relates the positive Lyapunov exponents of a map with its Kolmogorov-Sinai entropy
 (akin to a topological entropy) that for unimodal maps reads $h_{KS}=\lambda,\ \forall \lambda>0$, since $h$ can
 be understood as a proxy for $h_{KS}$. Unexpectedly, this qualitative agreement seems also valid in the periodic
 windows ($\lambda<0$), since the graph entropy is positive and approximately varies with the
 value of the associated (negative) Lyapunov exponent even though $h_{KS}=0$, hinting at a Pesin-like relation valid also out of chaos which deserves further
 investigation.
 The agreement between both quantities lead us to conclude that the Feigenbaum graphs capture not only the period-doubling route to chaos in a universal way, but also inherits the main feature of chaos, \textit{i.e.} sensitivity to initial conditions.\\

\section{Summary}

In summary, we have described how the horizontal visibility
algorithm transforms nonlinear dynamics into networks, in
particular how the entire Feigenbaum scenario manifests in network
language. The most important quality of the HV algorithm when
applied to this category of time series is that it leads to
definite and unique results, to well-defined families of graphs
and to analytic expressions for their main property, the degree
distribution. In obtaining these results two basic ingredients
played a decisive role; these are the predetermined order in which
positions or bands are visited in the time series and the
self-affinity of periodic or chaotic-band attractors. As we have
seen, the graphs generated by the algorithm store in the
connectivity pattern amongst nodes, i.e. in their topological
structure, the dynamical nature of the nonlinear map. A
significant feature of the Feigenbaum graphs is their evident
universality as the structure of all graphs and that of all
families they form is independent of the specific form of the
unimodal map including the degree of nonlinearity of its extremum.
We have also shown that the families of networks and degree
distributions obtained from periodic and chaotic attractor
bifurcation cascades have scale-invariant limiting forms. And that
the latter occupy the dominant positions of RG fixed points and
extrema of the entropy associated with the degree distribution.
The capability of the HV algorithm to expose new information is
indicated by the property of the network entropy to emulate the
Lyapunov exponent for both periodic and chaotic attractors.
Applications of this approach to other low-dimensional nonlinear
circumstances, such as the dynamical complexity associated with
the quasiperiodic route to chaos and the vanishing of Lyapunov
exponents are under study, and an obvious open question is how well would these precise analytical results
work for experimental data.

\noindent \textbf{Acknowledgments}\\
The authors acknowledge suggestions from anonymous referees.
BL and LL acknowledge financial support from the MEC and Comunidad
de Madrid (Spain) through projects FIS2009-13690 and
S2009ESP-1691. FJB acknowledges support from MEC through project
AYA2006-14056, and AR acknowledges support from MEC (Spain) and
CONACyT and DGAPA-UNAM (Mexican agencies).

%\bibliography{apssamp}% Produces the bibliography via BibTeX.

\newpage

\end{document}